\documentclass[aps,pra,twocolumn,superscriptaddress,english,showpacs]{revtex4-1}

\usepackage{natbib}
\usepackage[utf8]{inputenc}
\usepackage[T1]{fontenc}
\usepackage[english, french]{babel} 
\usepackage{amsmath}
\usepackage{amsfonts}
\usepackage{makeidx}
\usepackage{graphicx}
\usepackage{mathtools} 
\usepackage{braket} 
\usepackage[linkcolor=black, citecolor=black]{hyperref} 
\usepackage[usenames,dvipsnames]{xcolor} 

\usepackage{tikz} 
\usetikzlibrary{positioning}
\usepackage{ dsfont } 

\usepackage{standalone}

\usepackage{subcaption}

\newcommand{\define}{\ensuremath{ \overset{\text{def}}{=} }}

\renewcommand{\d}[1]{\mathrm{d}#1}

\newcommand{\simlim}[2]{\ensuremath{ \underset{#1 \rightarrow #2}{\sim} }}

\newcommand{\lb}{\ensuremath{\overline{\lambda}}}
\newcommand{\zb}{\ensuremath{\overline{z}}}
\newcommand{\wf}{\ensuremath{D^\psi}}
\newcommand{\avwf}{\ensuremath{\overline{D^\psi}}}

\newcommand{\avspec}{\ensuremath{D^\mu}}
\newcommand{\gv}{\ensuremath{\mathbf{g}}}
\newcommand{\sub}{\ensuremath{M}}

\begin{document}

\title{Fractal dimensions of the wavefunctions and local spectral measures on the Fibonacci chain.}
\author{Nicolas Macé}
\author{Anuradha Jagannathan}
\author{Frédéric Piéchon}
\affiliation{Laboratoire de physique des Solides, Université Paris-Saclay, 91400 Orsay, France}

\selectlanguage{english}

\date{\today}

\begin{abstract}
We present a theoretical framework for understanding the wavefunctions and spectrum of an extensively studied paradigm for quasiperiodic systems, namely the Fibonacci chain. Our analytical results, which are obtained in the limit of strong modulation of the hopping amplitudes, are in good agreement with published numerical data. In the perturbative limit, we show a new symmetry of wavefunctions under permutation of site and energy indices. 
We compute the wavefunction renormalization factors and from them deduce analytical expressions for the fractal exponents corresponding to individual wavefunctions, as well as their global averages. The multifractality of wavefunctions is seen to appear at next-to-leading order in $\rho$. 
Exponents for the local spectral density are given, in extremely good accord with numerical calculations.  Interestingly, our analytical results for exponents are observed to describe the system rather well even for values of $\rho$ well outside the domain of applicability of perturbation theory.  
\end{abstract}

\maketitle

\emph{Introduction}
 As distinct from periodic crystals on the one hand, where electronic
states are typically extended, and disordered systems on the other hand, where states are typically localized for low dimension and/or strong
enough disorder, electronic states in quasicrystals are believed to have an intermediate ``critical'' character.  The study of tight binding models on the Fibonacci chain, 
 a one dimensional paradigm for quasicrystalline structures, is particularly important, as a first step towards understanding the physics of these systems. These models have been extensively investigated theoretically, as in \cite{Kohmoto1983,tang1986global,Kohmoto1987,bellissard1989,damanik2009almost,MossSire}. There have also been many experimental studies of electronic properties of this model. To cite some recent works, in \cite{silberberg2015} investigated transport due to topologically protected edge states in a Fibonacci photonic waveguide. In \cite{tanese2014}, the  density of states of a Fibonacci tight-binding model was studied by direct observation of polariton modes in a one-dimensional cavity, and shown to have the fractal structure, log-periodic oscillations and gaps labeled as predicted by theory \cite{bellissard1989}. 

While the spectral properties in the model are now reasonably well understood, wavefunctions are less well characterized. More generally, despite the belief, supported by numerical evidence, that there are critical wave functions in quasicrystals, there are no analytical calculations for the fractal properties of all wavefunctions, to our knowledge. In view of the importance of the structure of eigenstates to understand, for example, dynamics or transport, obtaining a theoretical description of states in this 1D quasicrystal is necessary. In this paper, we present a detailed calculation of the properties of all the wavefunctions in the off-diagonal Fibonacci tight-binding model, in which the hopping terms have two possible amplitudes, ordered according to the 
Fibonacci sequence.  In the strong modulation regime where the ratio of hopping amplitudes $\rho \ll 1$, one can write an approximate renormalization group transformation for this model \cite{ KaluginKitaevLevitov,Niu1990, Piechon95} and obtain recursion relations for multifractal exponents of the spectrum.   Returning to this approach, we obtain explicit expressions for the fractal exponents corresponding to individual wavefunctions. We show, using the conumbering scheme, that wavefunctions are symmetric under exchange of site and energy indices in the perturbative limit. We show that the multifractal property of wavefunctions appears at next-to-leading order in $\rho$. We give expressions for their globally averaged values. We compute next the generalized exponents of the global and the local spectral measures. Our results agree very well with numerical computations on approximants of the Fibonacci chain. 

In Sec.\ref{sec:definitions} we introduce the model, and some basic definitions and notations. 
In Sec.\ref{sec:RG} we review the real space renormalization group used to calculate the wavefunctions in perturbation theory.
In Sec.\ref{sec:calculations1} we present the main steps of the calculation of fractal exponents, showing in detail how to obtain results to leading order. In the following section Sec.\ref{sec:calculations2}, we present the higher order corrections, and we obtain theoretical predictions which we then compare with numerical data.
In Sec.\ref{sec:discussion} we discuss some of the implications of these results. We present an inequality relating the different families of exponents which, we speculate, might be more generally valid in other 1D models. 
Summary and conclusions are presented in Sec.\ref{sec:conclusion}.
The Appendix shows details of the higher order calculation in perturbation theory. 


\section{\label{sec:definitions}Model and definitions}

The tight-binding Hamiltonian that we will consider in this work is the pure hopping model  given by 
\begin{eqnarray}
H=  \sum_i t_i \left( \ket{i} \bra{i+1} +  \ket{i+1} \bra{i} \right),
\end{eqnarray}
where the hopping amplitude between sites $i$ and $i+1$, $t_i$, can take the value $t_s$ (strong)  or $t_s$ (weak).
The ratio of the number of hopping amplitudes of each type is $N(t_w)/ N( t_s) = \omega$ where $\omega =2/(1+\sqrt{5}) $ is the inverse of the golden ratio. This ratio is irrational, showing that the chain cannot be periodic. 
$t_i$ varies along the chain according to the rule
\begin{eqnarray}
	t_i = \begin{cases}
	t_w & \text{when}~ i \bmod \omega^{-1} \geq \omega, \\
	t_s & \text{otherwise}.
	\end{cases}
	\label{condition}
\end{eqnarray}
The Fibonacci chain can alternatively be constructed recursively by the inflation method.
We start with the finite chain $C_0 = t_s$, to which we apply the inflation rule
\begin{equation}
	r \define \begin{cases}
        t_{w} &\rightarrow t_w t_s \\
        t_s &\rightarrow t_w
      \end{cases}
\end{equation} 
to obtain a series of longer and longer chains : $C_1 = r(C_0) = t_w$, $C_2 = r(C_1) = t_w t_s$, ... $C_n = r^n(C_0)$.
The Fibonacci chain is then defined as the semi-infinite chain $C_\infty$.

In numerical calculations, we will replace the Fibonacci chain $C_\infty$ by a periodic system, whose elementary cell is the finite chain $C_n$ (the so-called $n^\text{th}$ periodic approximant).
In the above formula, this amounts to replacing $\omega$ by a rational approximant, $\omega_n = F_{n-2}/F_{n-1}$, where $F_n$ is the $n^\text{th}$ Fibonacci number (starting from $F_1 = F_2 = 1$).

\begin{figure}[htp]
	\centering
	\includegraphics[scale=1]{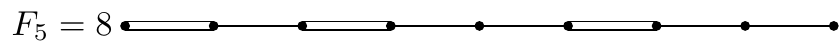}
	\caption{The periodically repeated block of the fifth approximant to the Fibonacci chain. Weak couplings $t_w$ are represented by a single line, and strong couplings $t_s$ by a double line.}
\label{fig:fib8}
\end{figure}

\emph{Atoms and molecules.} 
In the strongly modulated limit, one has a natural classification of sites into ``molecule-''type (m) and ``atom-''type sites (a) depending on their local environment. 
Atom-type sites have weak bonds on the left and on the right, so they are weakly coupled to the rest of the chain. 
Molecule-sites are linked by a strong bond to another molecule-site, and have a weak bond on either side. In the limit $t_w \rightarrow 0$, these pairs form 
isolated diatomic molecules, while the remaining sites correspond to isolated atoms.

\section{The perturbative renormalization scheme}
\label{sec:RG}

We now recall the main ideas behind the perturbative renormalization scheme introduced by Niu and Nori \cite{Niu1990}, and independently by Kalugin, Kitaev and Levitov \cite{KaluginKitaevLevitov}.
The Hamiltonian depends on the single parameter $\rho=t_w/t_s$, with $\rho \ll 1$ in the strong modulation limit.

When $\rho = 0$, the atoms and the molecules decouple. The spectrum consists of three degenerate levels: $E = \pm t_s$, corresponding to  molecular bonding and antibonding states, and $E=0$, for the isolated atomic state.
When $\rho \neq 0$, the states in each of the three degenerate levels weakly couples to each other, thus lifting the degeneracy.
We now consider separately the case of the atomic and of the molecular energy levels.

\begin{figure}[htp]
	\includegraphics[scale=1]{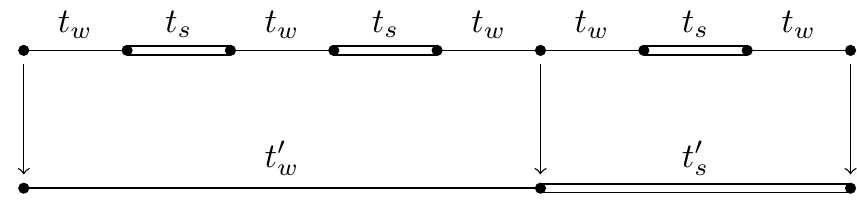}
	\caption{\small{Illustration of the atomic deflation rule: here the fifth approximant is transformed to the second.}}
\label{fig:at_defl}
\end{figure}

\emph{Atomic levels}
 At first order, each atomic energy level, localized on an atomic site couples to the atomic levels localized on the two neighboring atoms (figure \eqref{fig:at_defl}). In perturbation theory,
the effective bond coupling two neighboring atoms takes on only
 two possible values (as illustrated on figure \eqref{fig:at_defl}), a strong and a weak one, arranged again according to the Fibonacci sequence.
More precisely, upon replacement of the couplings between atoms by renormalized couplings, one passes from the chain $C_n$ to the chain $C_{n-3}$. 
We call this geometrical transformation the \emph{atomic deflation}.
The renormalized couplings are linked to the old ones by a multiplicative factor  $\zb$ \cite{Piechon95}.
One finds $t_s' = \zb t_s$, $t_w' = \zb t_w$,  with $\zb = \rho^2$.

\begin{figure}[htp]
	\includegraphics[scale=1]{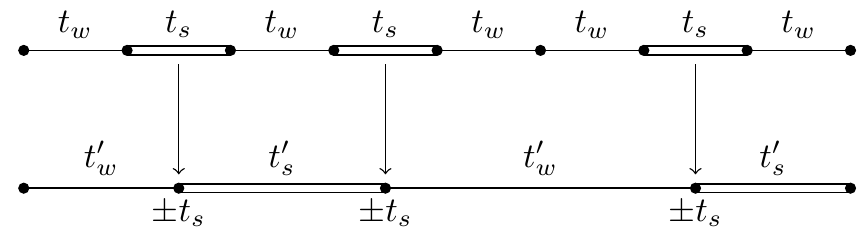}
	\caption{\small{Illustration of the molecular deflation rule: the fifth approximant is transformed to the third.}}
\label{fig:mol_defl}
\end{figure}

\emph{Molecular levels.}
In a similar fashion, each state localized on a diatomic molecule, is coupled to the neighboring molecules through only two possible effective couplings between two neighboring molecules.
Upon replacement of the couplings between molecules by renormalized couplings, one passes from the chain $C_n$ to the chain $C_{n-2}$ (see figure \eqref{fig:mol_defl} for an example).
We call this geometrical transformation the \emph{molecular deflation}.
The renormalized couplings are linked to the old ones by a multiplicative factor  $z$.
One finds $t_s'' = z t_s$, $t_w'' = z t_w$,  with $z = \rho/2$.
In addition, the molecular renormalization introduces on-site potentials $\pm t_s$ on the deflated chain, shifting the whole energy spectrum by $+ t_s$ (bonding molecular levels), or $- t_s$ (antibonding molecular levels).

To summarize, the Hamiltonian of the $n^\text{th}$ approximant decouples into the direct sum of three Hamiltonians:
\begin{equation}
\label{eq:recur_ham}
	H_n = \underbrace{\left( z H_{n-2} - t_s \right)}_{\text{bonding levels}} \oplus \underbrace{\left( \zb H_{n-3} \right)}_{\text{atomic levels}} \oplus \underbrace{\left( z H_{n-2} + t_s \right)}_{\text{antibonding levels}} + \mathcal{O}(\rho^4)
\end{equation}

In the limit $n \rightarrow \infty$, the chain becomes quasiperiodic and its wavefunctions and its spectrum become nontrivial: they exhibit multifractality \cite{Kohmoto1983,tang1986global,Kohmoto1987,Zheng1987}.

\subsection{Renormalization paths, equivalence between energy labels and conumbers.}

\begin{figure}[htp]
    \includegraphics[width=.4\textwidth]{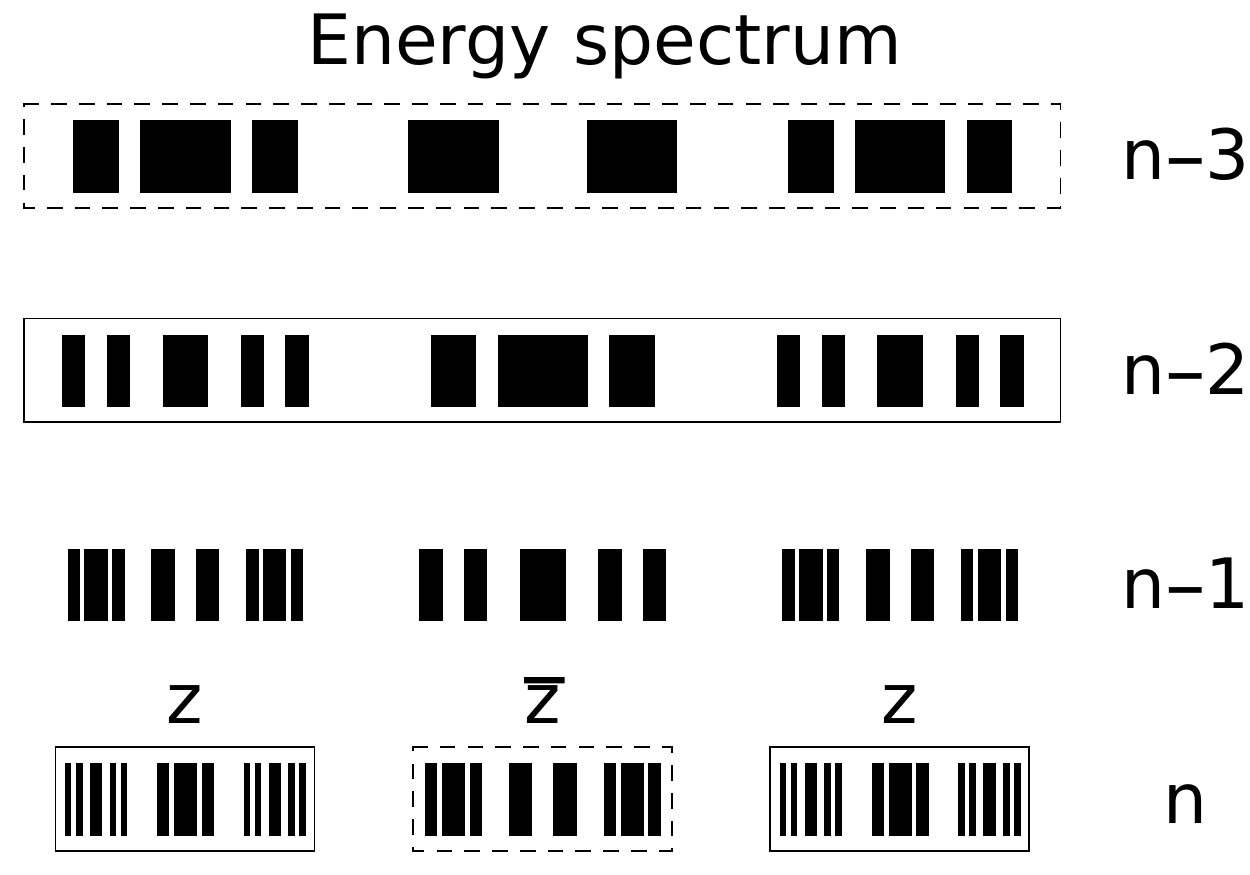}
\caption{\small {Spectrum of the approximant $H_n$ ($n = 8$) constructed geometrically from the spectra of $H_{n-2}$ and $H_{n-3}$ (relation \eqref{eq:recur_ham}). $z$ and $\zb$ are the two scaling factors.}}
\label{fig:recur_spec}
\end{figure}

\textbf{Renormalization paths of the energy bands.}\\
Eq. \eqref{eq:recur_ham} tells us that the spectrum of the Hamiltonian $H_n$ is the union of three energy clusters -- the antibonding molecular cluster, the atomic cluster and the bonding molecular cluster -- each of which is a scaled version of the spectrum of a smaller approximant.
Molecular clusters are separated from the atomic cluster by a gap of width $\Delta \sim t_s( 1 - z )$. Each of these main clusters can be decomposed into three sub-clusters, and so on.
The spectrum has therefore a recursive, Cantor set-like description, as shown by fig. \eqref{fig:recur_spec}.


\begin{figure}[htp]
    \includegraphics[width=.4\textwidth]{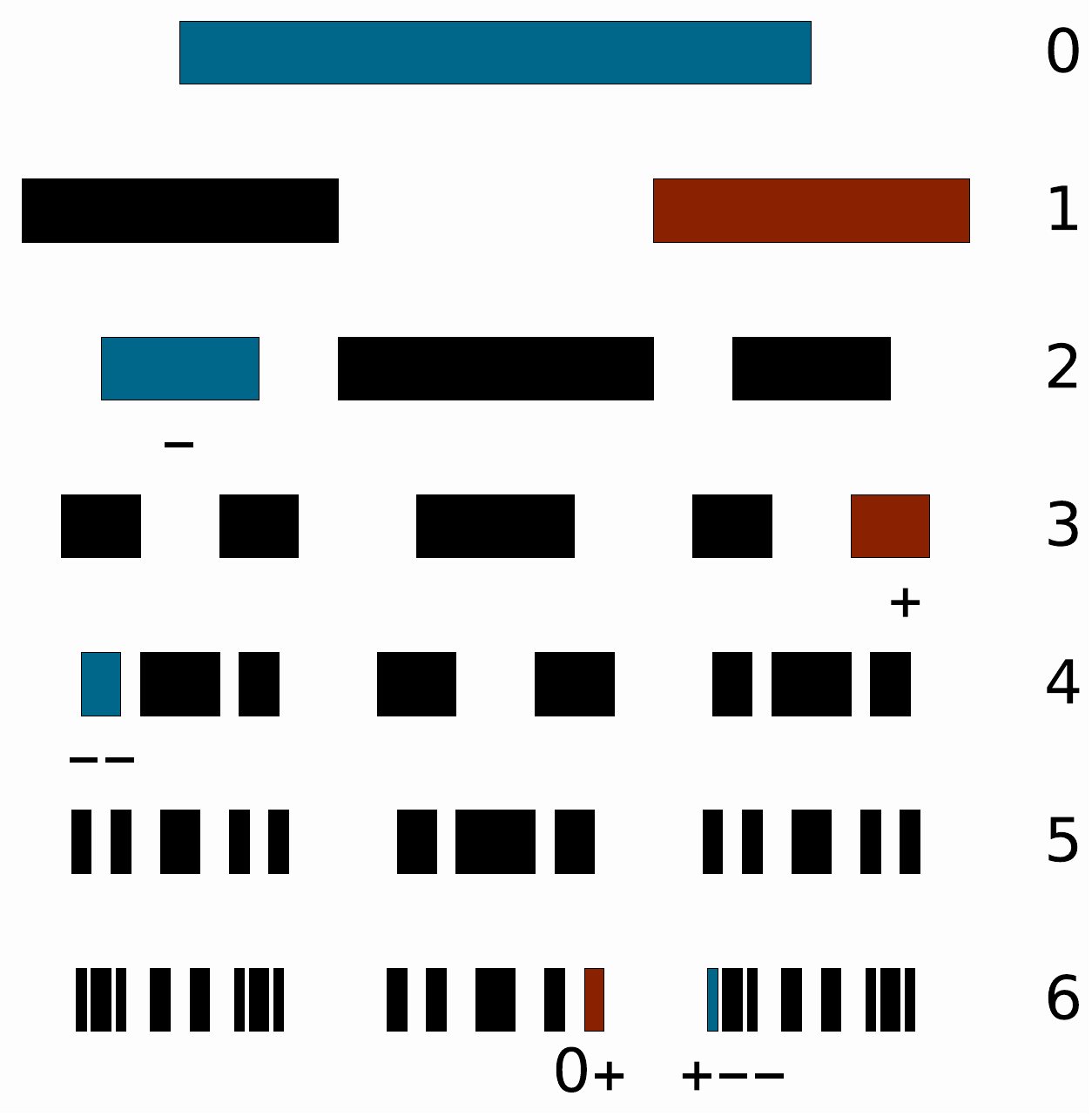}
\caption{\small {Every energy band can be labeled by its renormalization path. As an example, here are shown the renormalization paths of the energy bands labeled by $0+$ (red) and $+--$ (blue).}}
\label{fig:energy_path}
\end{figure}

One can assign to the energy bands belonging to the bonding, atomic or antibonding clusters respectively the labels $+$, $0$ or $-$.
To this, one can append another $+$, $0$ or $-$ according to the sub-cluster type of each energy. 
Repeating this procedure recursively, one obtains for each energy band a unique sequence of letters called its \emph{renormalization path} \cite{Piechon95, Kohmoto1987}. 
Figure \eqref{fig:energy_path} shows the renormalization paths of two particular energy bands.
Let us note that although the renormalization path labeling of the energy bands has been derived in the perturbative limit, it continues to hold for $0 < \rho < 1$ since no gap closes.

\textbf{Renormalization paths of the sites.}
One assigns to every atomic (resp molecular) site of the chain $C_n$ a label ``a'' (resp ``m''). 
Because the set of atoms and the set of molecules of $C_n$ are mapped to the chains $C_{n-3}$ and $C_{n-2}$ respectively, one can repeat the labeling procedure recursively.
Thus, to each site is associated a sequence of letters, that we call the \emph{renormalization path} of the site.
Note that because we have not distinguished between bonding and antibonding states on molecules, several molecular states can have the same renormalization path. 

\begin{figure}[htp]
    \includegraphics[width=.5\textwidth]{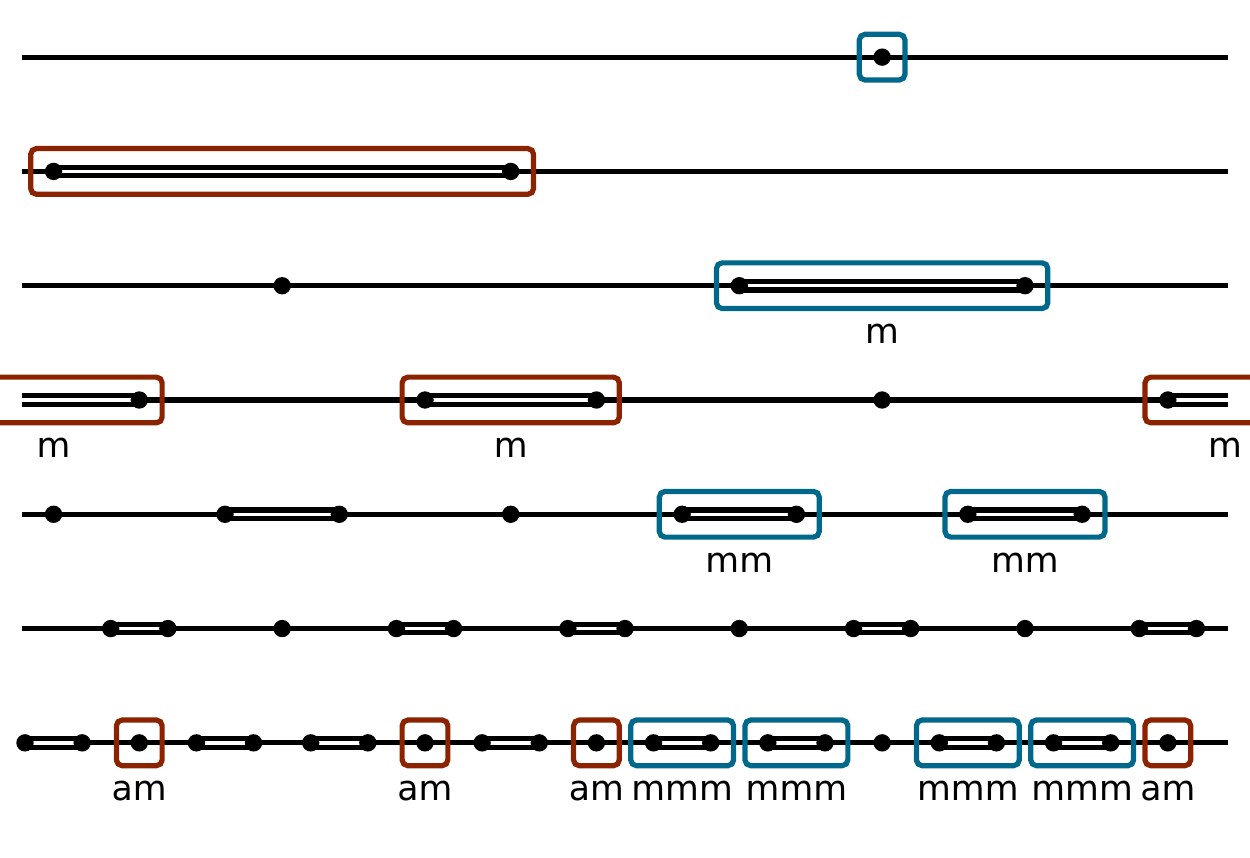}
\caption{\small {Every site can be labelled by its renormalization path, as shown here for the sites of non-zero amplitude, at first order, of the wavefunctions labelled by $0+$ (red) and $+--$ (blue).}}
\label{fig:chain_path}
\end{figure}

\textbf{Symmetry between renormalization paths for sites and energies.}
In the perturbative limit, because of \eqref{eq:recur_ham}, an eigenfunction associated to an atomic/molecular energy band has nonzero amplitude only on atomic/molecular sites. 
This is again true at every step of the renormalization process, so that by recursion, a given eigenfunction has nonzero amplitude at first order only on sites whose renormalization path matches the one of the energy level associated energy band (provided that we make the identification $\pm \leftrightarrow m$ and $0 \leftrightarrow a$).

\begin{figure*}
\centering
  \includegraphics[width=1.\textwidth]{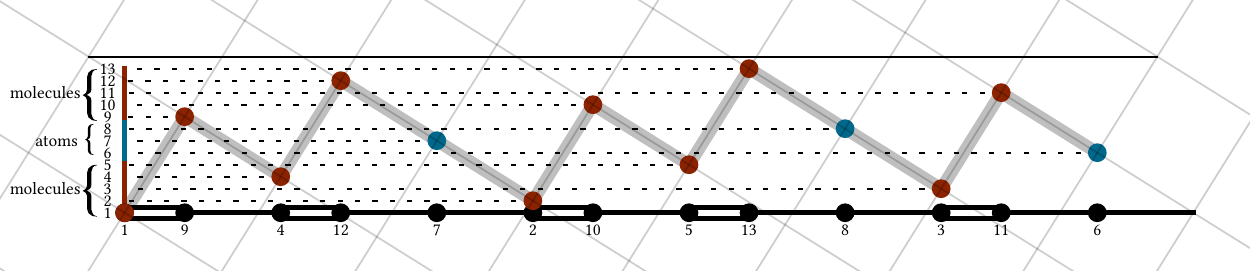}
\caption {\small {Example of the cut and project method showing sites along the horizontal physical axis, and their conumber along the perpendicular space (vertical) axis. Conumbering naturally orderes sites according to their local environment: atomic sites (in blue) are clustered around the center of the windows, while molecular sites (in red) are grouped at the sides.}}
\label{fig:cut_proj}
\end{figure*}

In order to further understand the symmetry between the renormalization paths for sites and energies, we now make use of the well known cut an project method.
Let us recall very briefly that the cut and project method considers the sites on the chain $C_n$ as projections along an axis $E_\parallel$ of selected sites on a square lattice (see fig. \eqref{fig:cut_proj}).
We can also consider the projection of the selected sites along the $E_\perp$ axis, orthogonal to $E_\parallel$.
The projection of the sites on $E_\perp$ forms a regular lattice whose density increases with the approximant size.
In $E_\perp$, as seen on fig. \eqref{fig:cut_proj}, the sites regroup in 3 clusters : a central atomic cluster surrounded by two molecular clusters.  
Keeping only sites belonging to the atomic/molecular clusters exactly amounts to performing an atomic/molecular decimation. 
Since the deflated chains are again Fibonacci chains, the 3 clusters are made of 3 molecular-atomic-molecular subclusters, and so on. 
Thus, in $E_\perp$ the sites are ordered exactly in the same way as the energy bands. This hints that looking at the sites in $E_\perp$ is of interest.
We therefore choose to label the sites according to their projection on $E_\perp$ (from bottom to top on fig. \eqref{fig:cut_proj}).
Let us call $i$ the index of each site according to the order in which it appears in real space.
Then, for the n$^\text{th}$ approximant, the conumbers $c$ are given by $c(i) = F_{n-1} i \mod F_n$.
This relabeling of the sites was first introduced by R. Mosseri \cite{Moss, MossSire}, and was called \emph{conumbering}.
Because of the symmetry between the ordering of the sites in $E_\perp$ and the ordering of the energy bands, the conumber labels play the same role as the energy labels. These symmetry will prove itself useful in our analysis of the wavefunctions.

\begin{figure}[htp]
\centering
  \includegraphics[width=.5\textwidth]{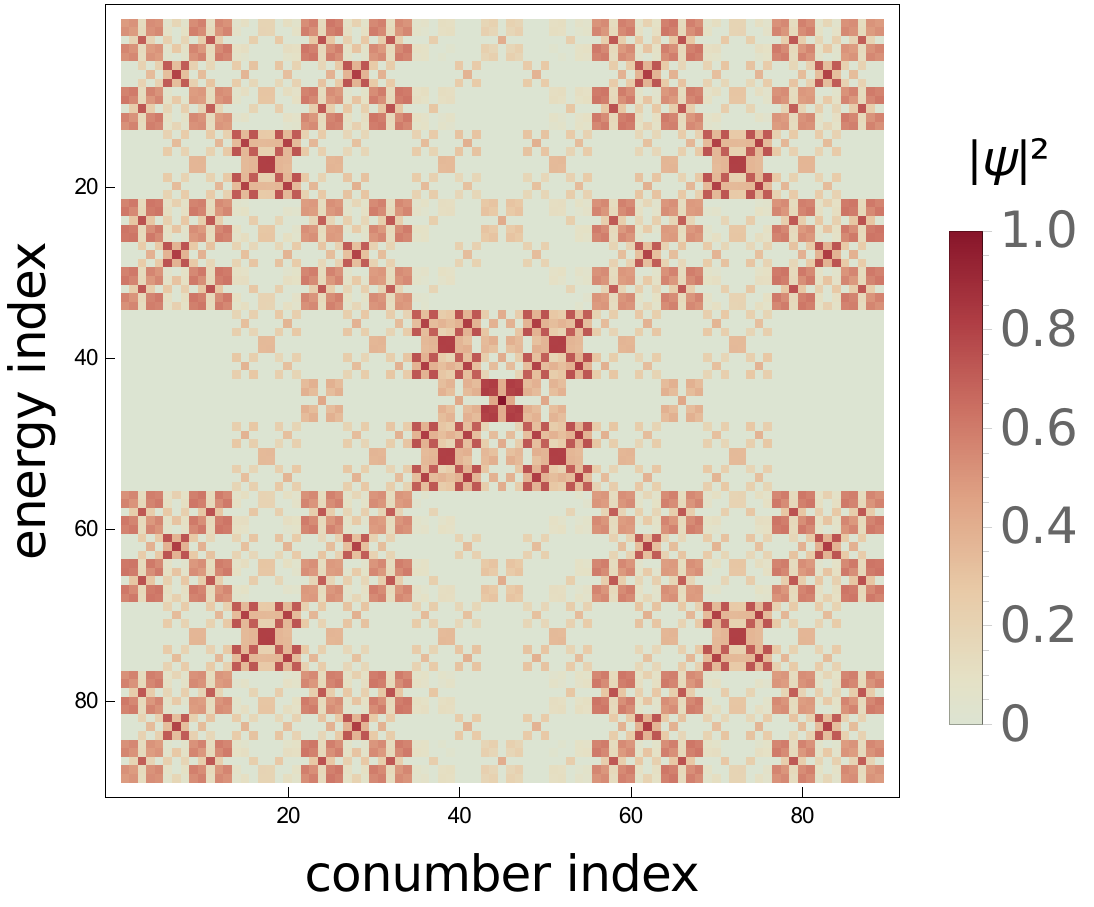}
  \includegraphics[width=.4\textwidth]{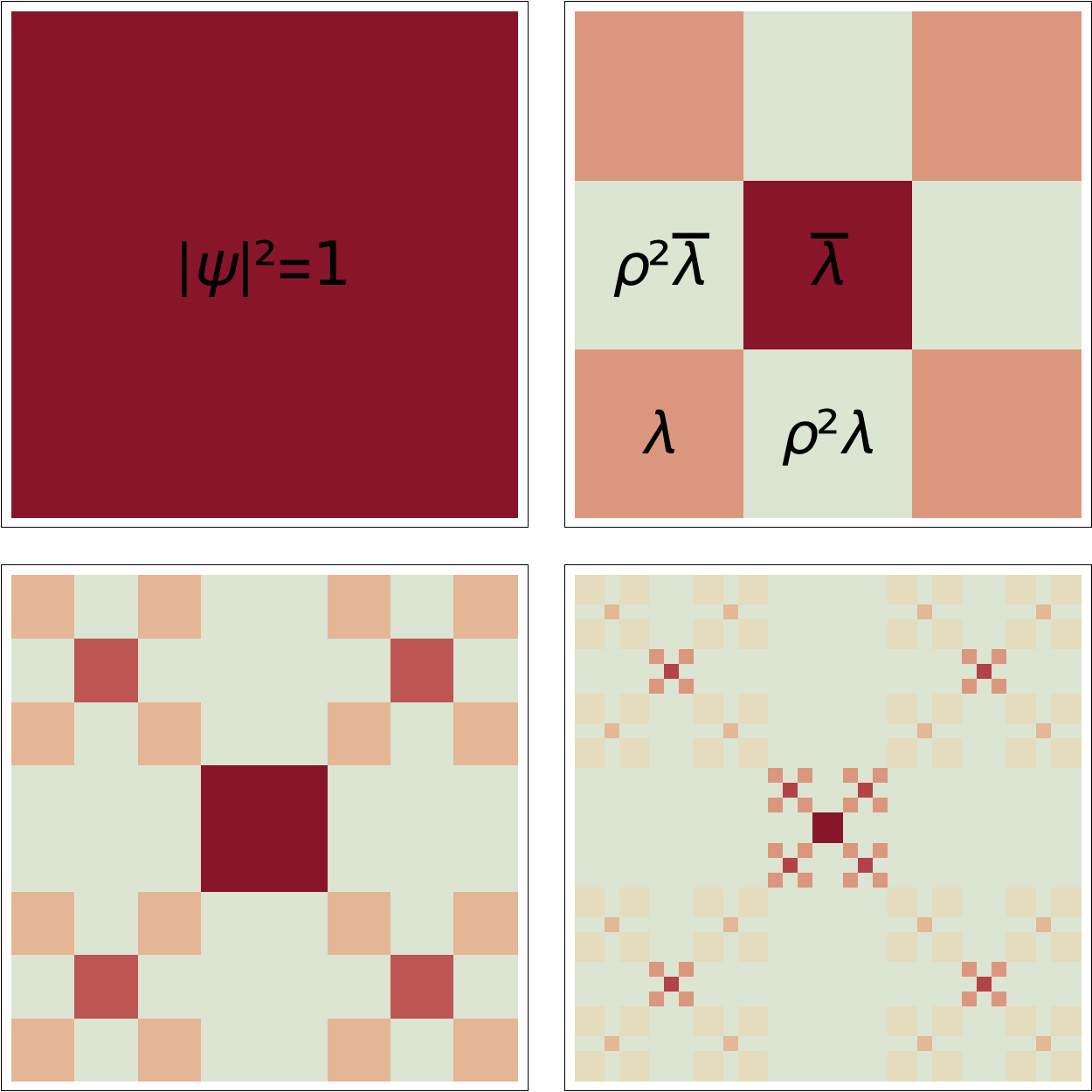}
\caption {\small {Upper figure: intensity plot of the numerically computed LDoS. x-axis: conumber index, y-axis: energy index. Color reprents the presence probability (ie the LDoS). Lower figure: the first few steps of the geometrical construction of the LDoS according to our perturbation theory.}}
\label{fig:wf_iDoS}
\end{figure}

As an illustration of the relevance of conumbering, we show the local density of states as a function of the energy labels and the conumber labels.
This plot is invariant under the exchange of the position/energy axes, in the $\rho \ll 1$ limit (figure \eqref{fig:wf_iDoS}), for the reason explained previously.
Crucially, this symmetry is revealed only if one uses conumbering for the sites, and remains hidden otherwise.

\subsection{The gap labelling theorem in the perturbative limit}
To conclude this section devoted to the spectral properties of the Fibonacci chain, we show how we can interpret the gap labelling theorem using the insight we gain from the renormalization scheme. 

In the gaps, the integrated density of states (IDOS) $N(E_{\text{gap}})$, takes values in a set specified by the gap labeling theorem.
In the case of the Fibonacci chain, the theorem states that

\begin{equation}
	N(E_{\text{gap}}) = \omega q \mod 1 = \omega q + p
\end{equation}
where $p$ and $q$ are integers. 
$q$ is the label of the gap, and $p(q)$ is such that the IDOS satisfies $0 \leq N \leq 1$.

\begin{figure}[htp]
\centering
  \includegraphics[width=.5\textwidth]{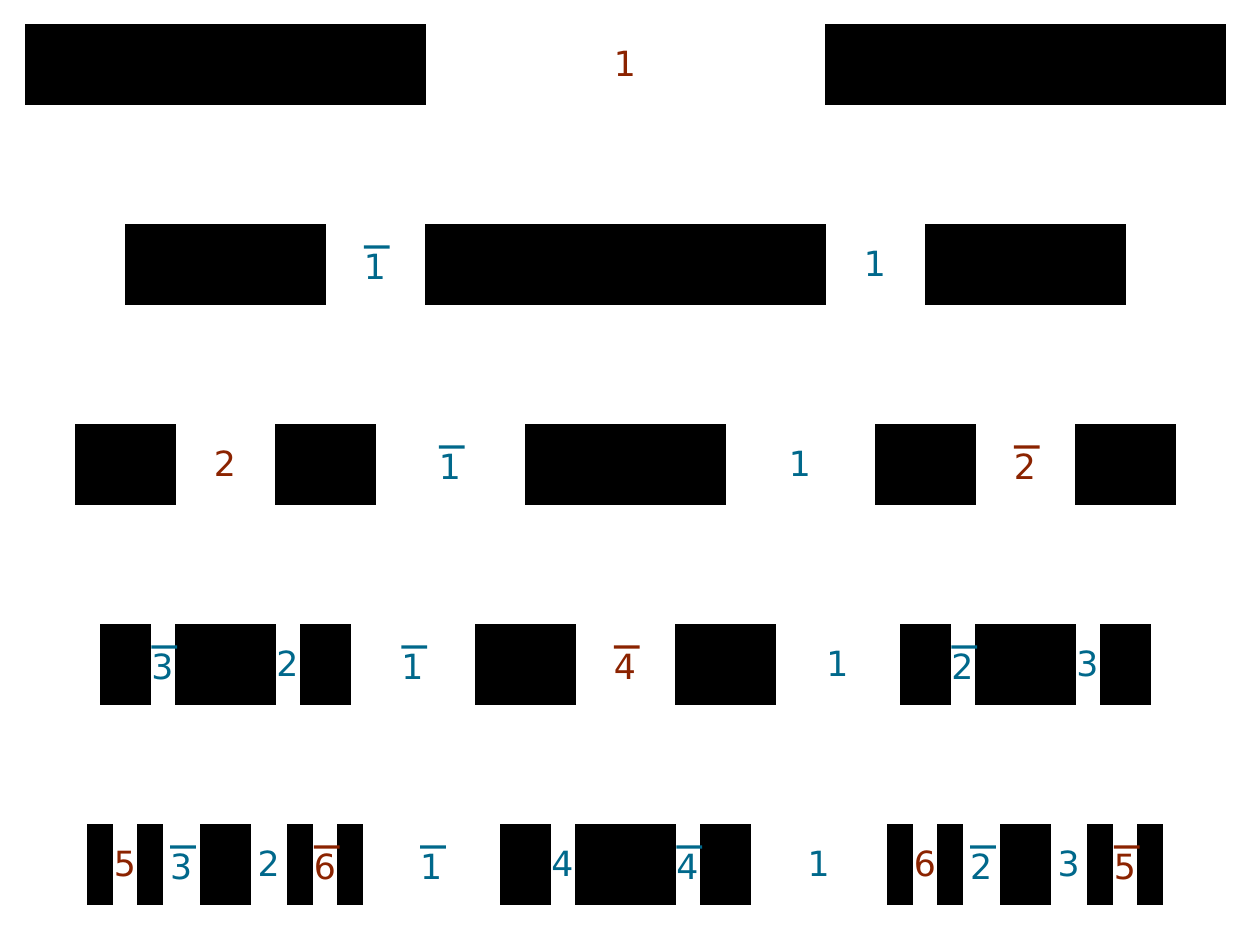}
\caption {\small {The gap labels, $q$, of the first few approximants. Blue: stable gaps, red: transient gaps. All the stable gaps are iterates of the 2 main gaps, while all the transient gaps are the iterates of the $E = 0$ gap.}}
\label{fig:gap_labels}
\end{figure}

For the n$^\text{th}$ approximant, the gap labelling becomes
\begin{equation}
	N_n(E_{\text{gap}}) = \omega_n q + p
\end{equation}
with $\omega_n = F_{n-1}/F_ n$, and $q \in [1, F_n)$.
As we have seen, in the strong modulation limit the spectrum has a hierarchical, ternary tree structure, and therefore the gaps have this structure as well.
To demonstrate the gap structure, let us call $G_n$ the set of IDOS of the gaps of the n$^\text{th}$ approximant. 
We then have the following recursion relations for the set of gap values in the bonding/atomic/antibonding clusters:
\begin{align}
\label{eq:recur}
	G_n^- &= \frac{F_{n-2}}{F_n} G_{n-2} \equiv d_-(G_{n-2})  \\
	G_n^0 & = \frac{F_{n-3}}{F_n}G_{n-3} + \frac{F_{n-2}}{F_n} \equiv d_0(G_{n-3}) \\
	G_n^+ & = \frac{F_{n-2}}{F_n} G_{n-2} + \frac{F_{n-1}}{F_n} \equiv d_+(G_{n-2})
\end{align}
where we have defined the three mappings $d-,~d_0,~d_+$ corresponding to the three different clusters.
These relations have a geometrical interpretation that can be best seen if we replace the gap $N_n(E_{\text{gap}}) = \omega_n q + p$ by the vector $\gv = (p, q)$. Then the above relations are replaced by the following affine transformations, depending on whether $\gv$ is in the bonding, atomic or antibonding cluster:
\begin{align}
	d_-(\gv) &= \sub^{-2} \gv \\
	d_0(\gv) &= \sub^{-3} \gv + \gv_1 \\
	d_+(\gv) &= \sub^{-2} \gv + \gv_2
\end{align}
where $\gv_1 = (1, -1)$ and $\gv_2 = (0, 1)$ are the labels of the two main gaps (corresponding to $q = \pm 1$, see figure \eqref{fig:gap_labels}),
and $\sub$ is the substitution matrix
\begin{equation}
	\sub = 
	\begin{bmatrix}
		1 & 1\\
		1 & 0\\
	\end{bmatrix}
\end{equation}
that generates the Fibonacci sequence by acting repetitively of the letters $A$ and $B$.

The recursive gap labeling procedure we just mentioned labels each of the $F_n -1 $ gaps of the n$^\text{th}$ approximant. However we know that some of these gaps are going to disappear in the quasiperiodic limit, while some are going to persist. Following \cite{Piechon95}, we call the former \emph{transient gaps} and the latter \emph{stable gaps}.
The gap at $E = 0$ that appears for even $F_n$ is an example of transient gap.
The two main gaps that separate the molecular clusters from the atomic cluster are examples of stable gaps.
From the recursive construction of the gap labels \eqref{eq:recur}, it is clear that the stable gaps are precisely the iterates through renormalization of the two main gaps, while the transient gaps are the iterates through renormalization of the $E = 0$ gap.

Let us stress that although the value of the IDOS inside a given \emph{stable} gap varies with the size of the approximant (but converges in the quasiperiodic limit), its gap label $(p,q)$ is \emph{independant of the size of the approximant}, and equal to the value we would have found for the infinite quasiperiodic system. 
In contrast, the gap label of a \emph{transient} gap varies with the size of the approximant. 
Also note that whereas the width of the transient gaps goes to zero in the quasiperiodic limit, their fraction stays finite. It converges to $(4 + 3 \omega)/(18+11\omega) \simeq 0.24$.
A more detailed discussion of the role of gap labeling for approximants is under preparation and will be given elsewhere \cite{Mace2016}.

To summarize this short section, we have seen that the renormalization group picture gives us a natural interpretation of the gap labelling theorem in terms of inflation/deflation transformations.
It has been shown that these gap labels can be physically interpreted in terms of the topological properties of edge states of a finite chain \cite{silberberg2015, Levy2015}.
We have also seen that the gap labelling, often mentioned in the context of quasiperiodic systems, naturally extends to periodic approximants. The only price to pay for this extension is the introduction of the transient and stable gaps.

\section{\label{sec:calculations1}Expressions for fractal dimensions to leading order}

We turn to the question of the fractal dimensions of the spectrum and of the wavefunctions in the limit $n \rightarrow \infty$.
For completeness, we first describe the derivation of the fractal dimensions of the spectrum \cite{Zheng1987,Piechon95}.

\subsection{Fractal dimensions of the spectrum }

The fractal dimensions of the global DoS can be determined using the thermodynamical formalism \cite{Halsey1986}.
 We define the partition function
 \begin{equation}
\label{eq:gamma_spec}
	\Gamma_n(q,\tau) = \sum_{E} \frac{\left( 1/F_n \right)^q}{(\Delta_n(E))^\tau}
\end{equation}
where $\Delta_n(E)$ is taken to be the width of the energy band associated to the energy level labelled $E$.
Separating the contributions of the bonding, antibonding and atomic energy levels, and using equation \eqref{eq:recur_ham}, we obtain
\begin{equation}
\Gamma_n = 2\left(\frac{F_{n-2}}{F_n}\right)^q z^{-\tau} \Gamma_{n-2} +  \left(\frac{F_{n-3}}{F_n}\right)^q \zb^{-\tau} \Gamma_{n-3}
\end{equation}
Taking the quasiperiodic limit, one obtains an implicit equation for the spectral fractal dimensions $D_q$:
\begin{equation}
\label{eq:spec}
	2 \omega^{2 q} z^{-(q-1)D_q}+\omega^{3 q} \zb^{-(q-1)D_q} = 1.
\end{equation}

We can solve it at first order in $\rho$, obtaining
\begin{equation}
	D_q = \frac{1}{1-q} \frac{\log \big[\omega^{-q} \left( \sqrt{1+\omega^{-q}} -1\right) \big]}{\log \rho} + \mathcal{O}\left( \frac{1}{(\log \rho)^2} \right)
\end{equation}

For the case $q=0$, we recall, one obtains the Hausdorff dimension $D_0$, which helps to characterize the nature of the spectrum. 
$D_0 = 0$ for a pure-point spectrum, while $D_0 = 1$ indicates that the spectrum has an absolutely continuous component.
An intermediate value $0 < D_0 < 1$ is the signature of a fractal spectrum.  Multifractality corresponds to situations where $D_q$ varies with $q$, which is the case here.

Here, we find
\begin{equation}
	D_0 = \frac{\log( \sqrt{2} -1 )}{\log \rho} + \mathcal{O}\left( \frac{1}{(\log \rho)^2} \right)
\end{equation}
in agreement with the result of Damanik \& Gorodetski \cite{DamanikGorodetski}, using trace-map-based methods.
For $\rho > 0$, one sees that $0<D_0<1$, and 
therefore we recover the well-established result that the spectrum of the Fibonacci Hamiltonian is fractal for nonzero $\rho$, however small.

These results compare well with numerical data only for extremely small values of $\rho$. 
Note however that improved multifractal analysis that goes beyond the first order was carried out \cite{Rudinger1996}.

\subsection{Fractal dimensions of the wavefunctions}
The fractal dimensions $\wf_q(E)$ of the wavefunction associated to the energy $E$ are defined by
\begin{equation}
	\chi_q^n(E) = \sum_i |\psi_i^n(E)|^{2q} \simlim{n}{\infty} \left( \frac{1}{F_n} \right)^{(q-1)\wf_q(E)}
\end{equation}
$\chi_2(E)$ is the inverse participation ratio, and the exponent $\wf_2(E)$ provides information as to the degree of localization of the state.
The value $D^\psi_2(E) = 1$ indicates that the state $E$ is extended, while $D^\psi_2(E) = 0$ characterizes a localized state.
Intermediate values $0 < D^\psi_2(E) < 1$ are a signature of a critical state, whose multifractal properties can be probed by varying $q$.

\begin{figure}[htp]
\centering
\includegraphics[width=.4\textwidth]{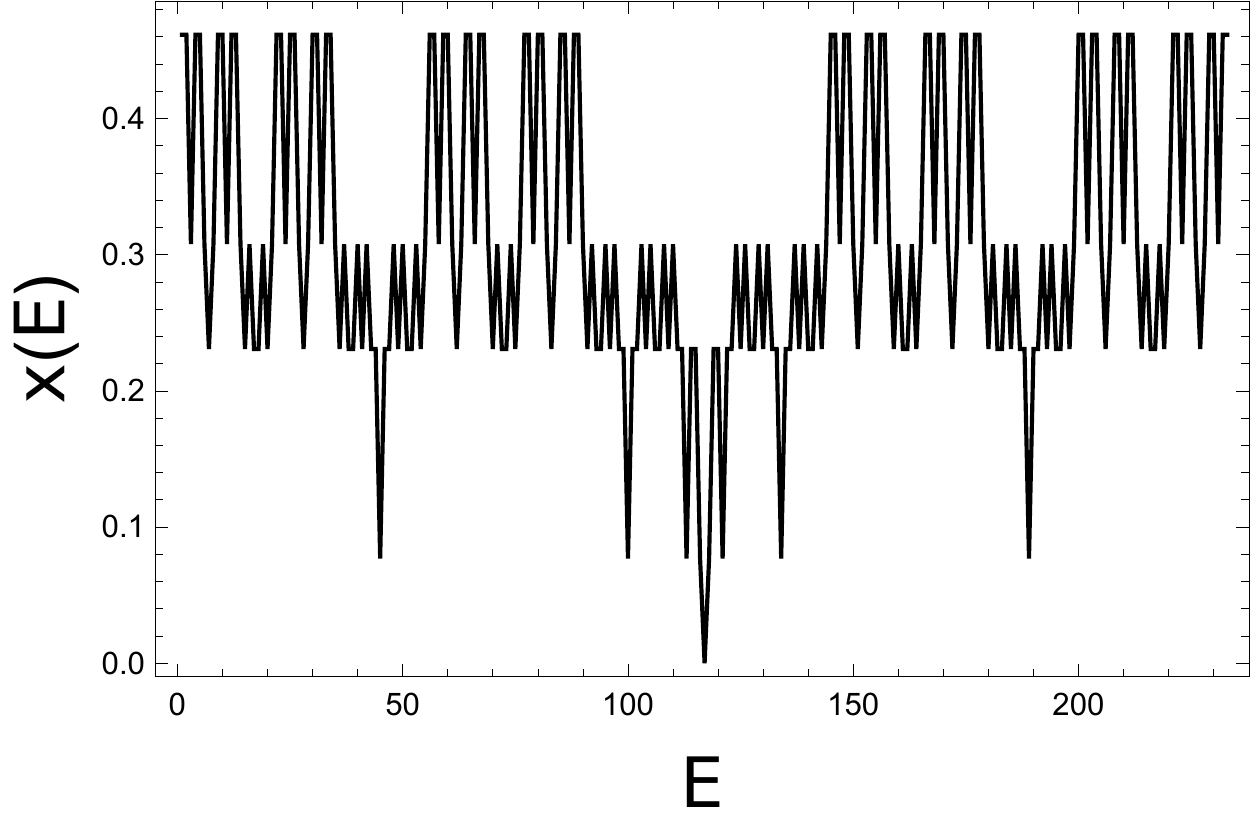}
\caption{\small{The parameter $x$ as a function of the energy labels (or equivalently of the sites in conumbering), for the approximant of 233 sites. Lines are drawn to guide the eye.}}
\label{fig:x}
\end{figure}

At leading order in $\rho$ the renormalization of the eigenstates is simple, and well understood \cite{Piechon96}.
For a wavefunction in the atomic cluster, we have $|\psi_i^n(E)| = |\psi_{i'}^{n-3}(E')| $.
For a wavefunction in the molecular cluster, $|\psi_i^n(E)| = |\psi_{i'}^{n-2}(E')|/\sqrt{2}$. $E$ and $E'$ are the energies on the original chain and on the deflated one respectively.
Therefore, we obtain immediately the leading order fractal dimensions of the wavefunction associated to the energy $E$:
\begin{equation}
\label{eq:dqpsi0}
	\wf_{q,0}(E) =- x(E) \frac{\log 2}{\log \omega} + \mathcal{O}(\rho^2),
\end{equation}
where
\begin{equation}
\label{eq:x}
	x(E) = \lim_{n \rightarrow \infty} \frac{n_m(E)}{n}
\end{equation}
with $n_m(E)$ the number of $+$/$-$ letters in the renormalization path of $E$, i.e. $x(E)$ is the fraction of RG steps spent in molecular clusters. 
$x$ is a non trivial function of the energy (figure  \eqref{fig:x}), whose structure is reminiscent of the one of the local density of states \eqref{fig:wf_iDoS}.
In the quasiperiodic limit, $x$ varies continuously between 0 and $1/2$, and $\wf_{q,0}(E)$ is a continuous function of $x$.
The distribution of $x$ is given by
\begin{equation}
	\Omega(x(n_0,n_m)) = 2^{n_m}\frac{(n_0+n_m)!}{n_0!n_m!}
\end{equation}
Where $\Omega(x)$ is the number occurrences of a given value of $x$.
It coincides with the distribution of the widths of the energy bands \cite{Piechon95}. Therefore, in the quasiperiodic limit, the distribution of $x$ is given by $f$, the Legendre transform of the fractal dimensions of the spectrum: $P(x) \sim F_n^{f(x)-1}$.
This distribution is sharply peaked around the most probable value, $x_{mp} = 2(3 \omega -1)/5 \simeq 0.3416...$.
States with this value of $x$ are statistically the most significant.

Since the fractal dimensions only depend on $x$, we perform the change of variables $\wf_{q,0}(E) \rightarrow \wf_{q,0}(x)$.
Since $0 \leq x \leq 1/2$, we have $0 < D^\psi_q(x) < 1$: the wavefunctions are critical, as we expect for a quasiperiodic system.

The parameter $x$ determines completely the fractal properties of the wavefunctions. 
For example, $x_a = 0$ for the level in the center having the renormalization path $00...$ (cf figure \eqref{fig:x}). The corresponding eigenstate has a zero fractal dimension, and is thus completely localized. 
On the other hand, the maximal value of $x=\frac{1}{2}$ is reached for the levels $E=E_\text{min}, E_\text{max} = \pm t_s/(1-z)$ at the edges of the spectrum, for which the renormalization paths are $++...$ and $---..$.
The corresponding eigenstates are the most extended. They occupy a fraction $(1/F_n)^{-\frac{\log 2}{2 \log \omega}}$ of the sites. 

To conclude, we note that, to leading order in $\rho$, the fractal dimensions of the wavefunction do not depend on $q$. 
Thus the wavefunctions are not multifractal at this order in $\rho$ and multifractality appears only at the next-to-leading order, as discussed in the next section. This first-order description of the wavefunctions has been compared to numerical results in \cite{Thiem2013}, where the agreement was found not to be very good. We argue that this is because the wavefunctions becomes rapidly multifractal as $\rho$ is increased.
Fortunately it is possible to calculate higher order corrections, and thereby vastly improve our theoretical predictions concerning the wavefunctions, as shown below.

\section{\label{sec:calculations2}Higher order renormalization group and multifractality}
 
At higher order the picture of molecular and atomic eigenstates and energies remains relevant, but it is now possible for an atomic eigenstate to have nonzero amplitude on molecular sites, and vice-versa.
In this section we explain our ansatz for the wavefunctions. In the next section, we will apply it to the computation of the fractal dimensions of the wavefunctions.

\subsection{Renormalization group for the wavefunctions}

\begin{figure}[htp]
	\includegraphics[scale=1]{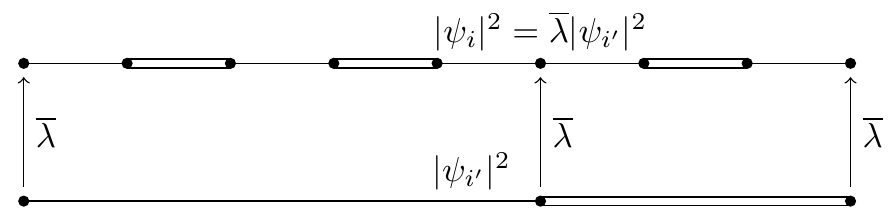}
	\includegraphics[scale=1]{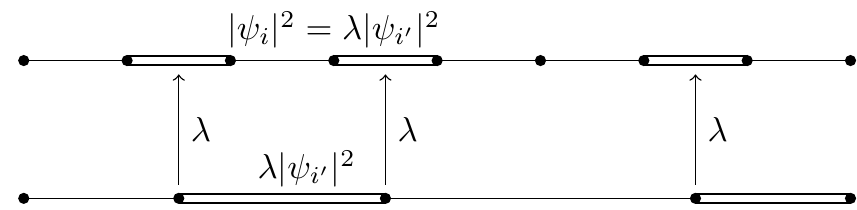}	\caption{\small{Schematization of the RG procedure. Top figure: a wavefunction of the atomic cluster, bottom figure: a wavefunction of the molecular cluster.}}
\label{fig:rg_at_wf}
\end{figure}

At leading order in $\rho$, we know that the wavefunction amplitudes on the $n^\text{th}$ approximant are related by a trivial multiplicative factor to the ones on a smaller approximant. 
At higher order, we still relate wavefunction coefficients on large approximants to wavefunctions on smaller ones through multiplicative factors.
We call these multiplicative factors $\lb$ if the wavefunction is of atomic type, and $\lambda$ is it is of molecular type (figure \eqref{fig:rg_at_wf}).
\begin{equation}
		\begin{cases}
		|\psi_i^{(n)}(E)|^2 = \lb |\psi_{i'}^{(n-3)}(E')|^2 \text{~if $E$ is atomic}\\
		|\psi_i^{(n)}(E)|^2 = \lambda |\psi_{i'}^{(n-2)}(E')|^2 \text{~if }E\text{~is molecular}
	\end{cases}
\end{equation}

$\lambda$ and $\lb$ are renormalization group parameters, and they play in the renormalization of the wavefunctions the role $z$ and $\zb$ plays in the renormalization  of the energy bands.

We find (details of the calculations are given in the Appendix):
\begin{align}
\label{eq:lb}
	\lb(\rho) &= \frac{2}{(1+\rho^2)^2 + \sqrt{(1+\rho^2)^4+4\rho^4}} \\
	\lambda(\rho) &= \frac{1}{1+\rho^2 \gamma(\rho) + \sqrt{1+(\rho^2 \gamma(\rho))^2}} 
\end{align}
with $\gamma(\rho) = 1/(1+\rho^2)$. 
At leading order in $\rho$, we recover $\lambda(0) = 1/2,~\lb(0) = 1$ as expected.
At next order,
\begin{align}
	\lb(\rho) = \frac{1}{1+\rho^2} + \mathcal{O}(\rho^2) \\
	\lambda(\rho) = \frac{1}{2+\rho^2} + \mathcal{O}(\rho^2)
\end{align}
Although our calculations are done in the strong modulation limit, in the periodic limit $\rho \rightarrow 1$, we obtain the exact expression for the renormalization factors: 
\begin{align}
	\lb(\rho) &\xrightarrow[\rho \to 1]{} \omega^3 \\
	\lambda(\rho) &\xrightarrow[\rho \to 1]{}  \omega^2
\end{align}.

\subsection{Local wavefunction dimensions}

For $q \geq 0$ and when $\rho \ll 1$,  we can write
\begin{equation}
	\chi_q^n(E) \simeq \begin{cases*}
	\left( \lb(\rho)^q / \lb(\rho^q) \right)\chi_q^{n-3}(E') & if $E$ is atomic, \\
	\left( \lambda(\rho)^q / \lambda(\rho^q) \right)\chi_q^{n-2}(E') & if $E$ is molecular. \\
	\end{cases*}
\end{equation}
Iterating this relation, we understand that the fractal dimensions depends only on the renormalization path of the energy $E$ we started with. 
Actually, it only depends on the parameter $x$ \eqref{eq:x}. 
Solving the recurrence we obtain an explicit expression for the fractal dimensions in the quasiperiodic limit:
\begin{equation}
\label{eq:dqpsi2}
	(q-1)\wf_q(x) =\log \left[  \left( \frac{\lambda(\rho)^q}{\lambda(\rho^q)} \right)^x \left( \frac{\lb(\rho)^q}{\lb(\rho^q)} \right)^{(1-2x)/3} \right]/\log \omega.
\end{equation}
It is easy to check that we recover the first-order expression for the fractal dimensions \eqref{eq:dqpsi0} if we take $\rho = 0$.

\begin{figure}[htp]
\centering
  \includegraphics[width=.45\textwidth]{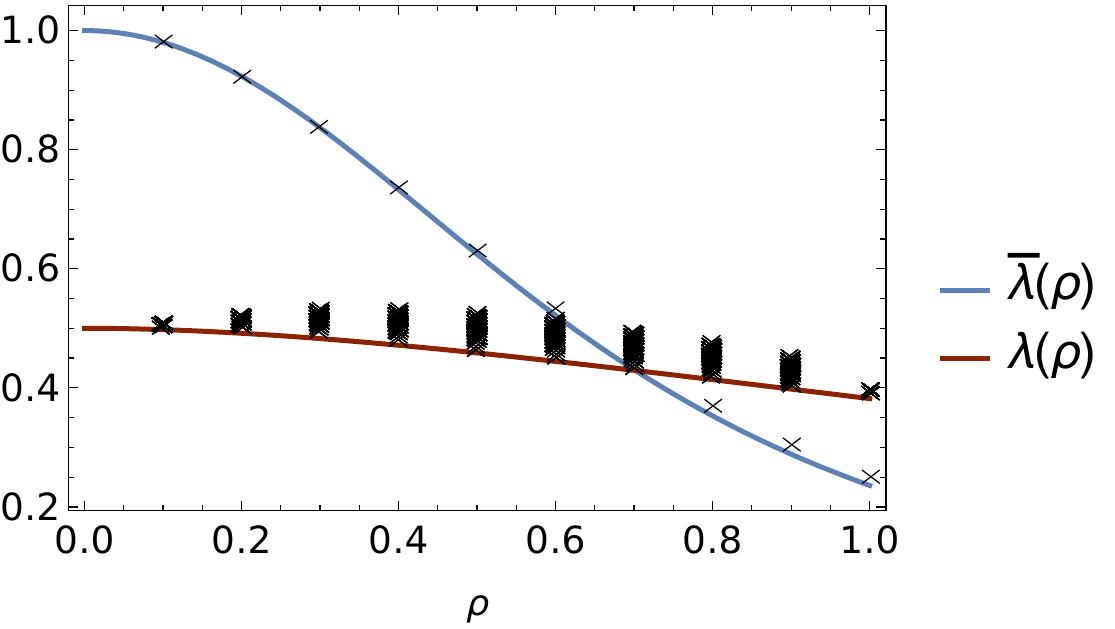}
  \caption{\small{Numerical results and theoretical predictions for the renormalization factors $\lb(\rho)$ and $\lambda(\rho)$.
  Dots: Numerical results ($n=19$, 4181 sites). Solid lines: theoretical predictions \eqref{eq:lamb}.
 }}
\label{fig:lambda_num}
\end{figure}

From this we can express the renormalization factors in terms of the fractal dimensions for $q \to \infty$, setting $x=0$ for $\lb$, and $x = 1/2$ for $\lambda$:
\begin{align}
\label{eq:lamb}
	\lb(\rho) &= \omega^{3 \wf_{\infty}(0)} \nonumber \\
	\lambda(\rho) &= \omega^{2 \wf_{\infty}\left(\frac{1}{2}\right)}
\end{align}
Thus, computing numerically the fractal dimensions for $q$ large gives us a numerical estimation of the renormalization factors, that we can compare to the theoretical predictions.
We expect the agreement to be good in the strong modulation $\rho \ll 1$ and in the weak modulation $\rho \sim 1$ regimes, because we know that in these limits the renormalization factors are exact.
In fact, we see that the agreement with numerics is excellent for all values of $\rho$ (fig .\eqref{fig:lambda_num}). 

It can be seen in fig \eqref{fig:lambda_num} that the values of $\lambda$ calculated numerically have a spread, which is not described by the theoretical formula. 
This is because, when $\rho$ is far from 0 or 1, $x$ is not enough to describe the renormalization parameter $\lambda$. One actually needs to know the whole renomalization path of the states, whereas our formula takes into account only  a single parameter $x$. 
The theoretical expression gives $\lambda$ accurately in only two cases: the states that are always bonding or always antibonding (ie the states at the edge of the spectrum). 
The states with alternating bonding and antibonding character have the largest deviation from the theoretical value, in the perturbative limit. 
The figure shows, as well, that the agreement with numerics is extremely good for the $\lb$ renormalization factor, corresponding to the wavefunction at $E = 0$. The small discrepancy between the numerics and the analytical predictions is only due to numerical finite-size effects.
Indeed, the fractal dimensions of this $E=0$ wavefunction have been determined  exactly \cite{Kohmoto1987}, using trace map methods.
Interestingly, these exact fractal dimensions coincide with our perturbative predictions for $\lb$ \eqref{eq:lamb}, for all values of $\rho$, meaning that our perturbative expression for $\lb$ is in fact exact.


\begin{figure}[htp]
\centering
  \includegraphics[width=.45\textwidth]{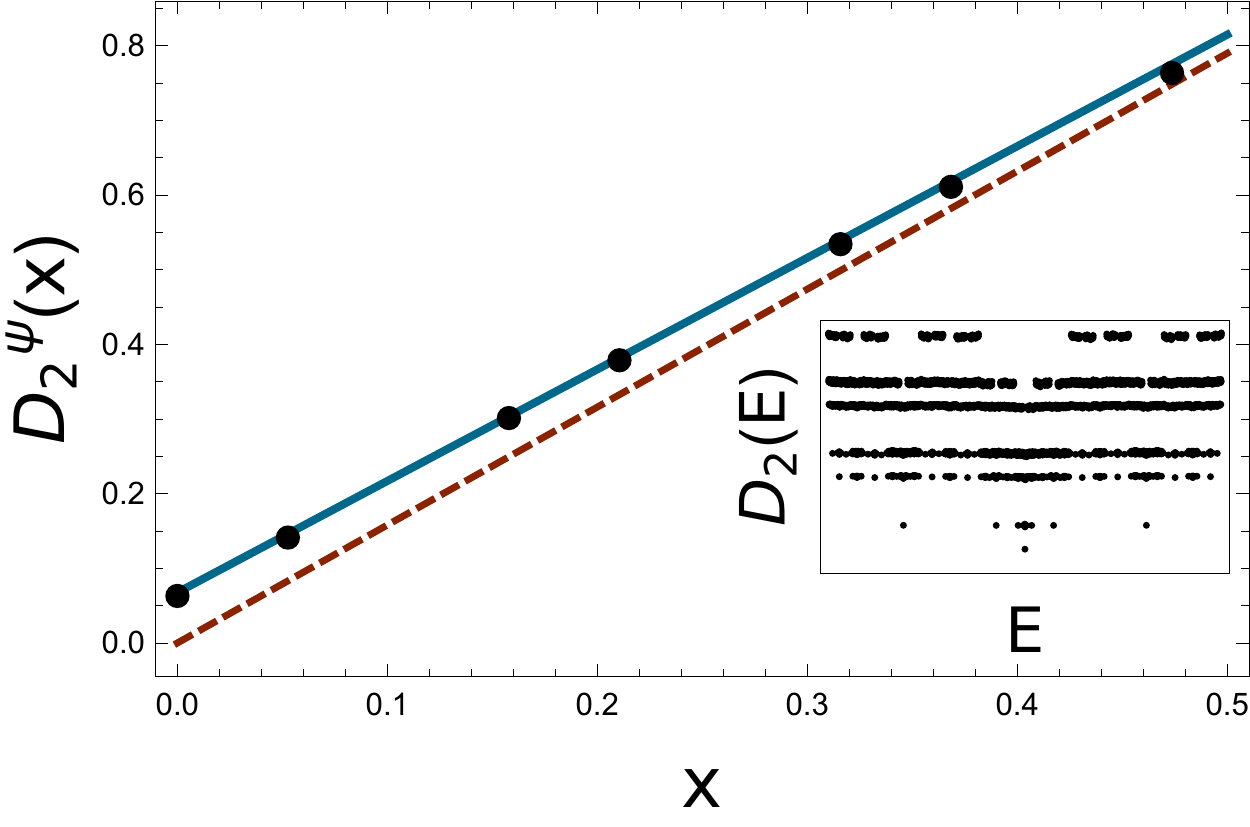}
  \caption{\small{Numerical results and theoretical predictions for the fractal dimensions of the wavefunctions.
  Dots: Numerical results ( $n=19$, 4181 sites). Dashed line: theoretical prediction at leading order (eq. \eqref{eq:dqpsi0}), solid line: theoretical prediction including multifractal corrections (eq. \eqref{eq:dqpsi2}).
  Inset: The fractal dimension $\wf_2(E)$ for every energy. In accordance with theoretical predictions, the fractal dimensions organize in lines, each line corresponding to a given value of $x(E)$.}}
\label{fig:dqpsi}
\end{figure}

We now check our theoretical predictions against numerical results.
The inset of figure \eqref{fig:dqpsi} shows how the fractal dimension for a given $q$ (here we chose $q=2$) depend on the chosen state of energy $E$.
We observe that the value of the fractal dimension organizes in lines. Along each line $\wf_q(E)$ is constant up to small variations and corresponds to a given value $x(E) = x$. 
That is, up to small corrections that should vanish in the limit $\rho \rightarrow 0$, the fractal dimension of a given wavefunction does not depend on the energy $E$, but only on $x(E)$. This is in agreement with the theoretical predictions \eqref{eq:dqpsi2}.
Figure \eqref{fig:dqpsi} shows, for example, the $x$ dependence of the fractal dimension for $q=2$. 
Calculated numerical results are seen to be in very good agreement with the theoretical predictions.
As the figure shows, the multifractal properties of the wavefunctions -- which were not captured at leading order in $\rho$ -- are relevant, even for the small value coupling ratio $\rho=0.1$. 
For $x=0$ the first order contribution to the fractal dimensions vanishes, so that only the multifractal correction term remains.

\begin{figure}[htp]
  \centering
  \includegraphics[width=.45\textwidth]{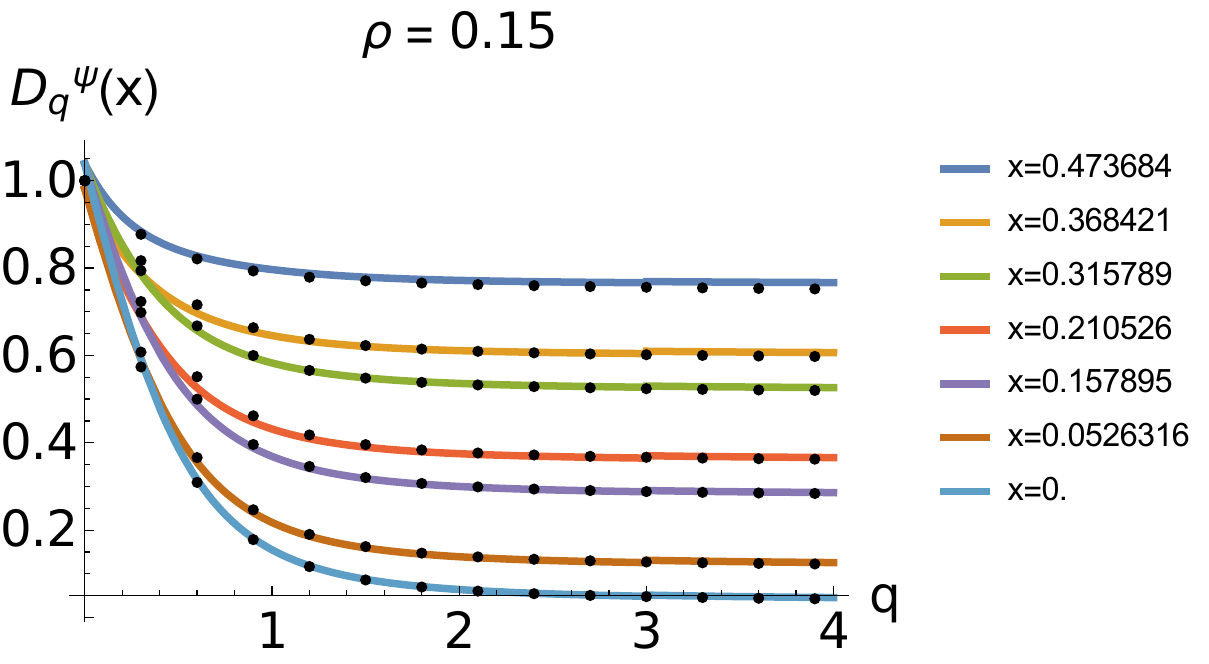}
  \caption{\small{The fractal dimensions $\wf_q(x)$ of the wavefunctions for the different values of $x$ accessible numerically. Dots are numerically computed data points, solid lines are the theoretical predictions (eq. \eqref{eq:dqpsi2}).}}
\label{fig:dqpsiq}
\end{figure}

To conclude this section, we show the $q$ dependance of the fractal dimensions for fixed values of $x$ (fig. \eqref{fig:dqpsiq}). 
The agreement with the theoretical predictions is excellent for all positive values of $q$.
This demonstrates that our theoretical analysis indeed captures the $q$ dependance of the fractal dimensions.
Since it is the multifractality of the wavefunctions that is responsible for the nontrivial $q$ dependance of the fractal dimensions, we conclude again that the multifractal corrections are relevant even at small coupling.

\subsection{The spectrally averaged fractal dimensions of wavefunctions}

\begin{figure}[htp]
	\centering
	\includegraphics[width=.45\textwidth]{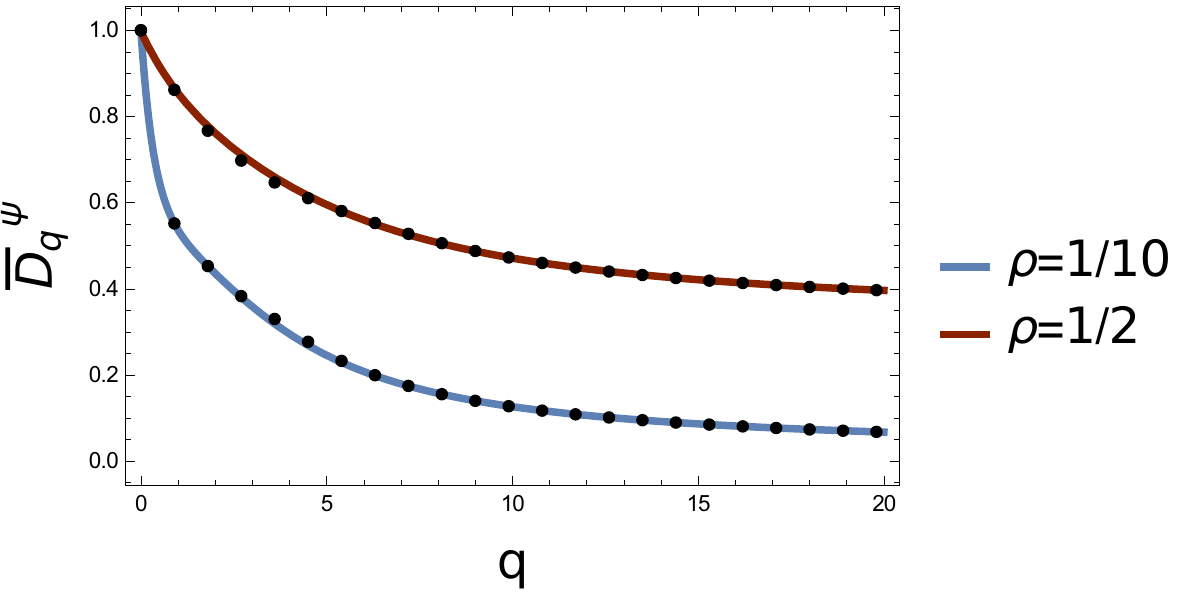}
	\caption{\small{The averaged fractal dimensions of the wavefunctions $\avwf_q$ as a function of the multifractal parameter $q$, for $\rho = 0.1, 0.5$. Dots: numerical results, solid line: theoretical predictions.}}
	\label{fig:avwf}
\end{figure}

In the previous section we have defined the fractal dimensions of individual wavefunctions, and have seen that they were associated with the energy level parameter $x$.
In this section we calculate scaling properties of the wavefunctions after averaging over all states.  We define an averaged fractal dimension $\avwf_q$ by
\begin{equation}
	\langle \chi_q \rangle = \frac{1}{F_n} \sum_{E} \chi_q(E) \sim \left( \frac{1}{F_n} \right)^{(q-1)\avwf_q}
\end{equation}
This quantity, studied in disordered systems at the Anderson localization transition \cite{Mirlin2006}, has been computed numerically by \cite{Thiem2013} for the Fibonacci model.
Within our perturbation theory, we obtain an implicit equation for the averaged fractal dimensions:
\begin{equation}
\label{eq:wf_av}
	2 \omega^2 \frac{\lambda(\rho)^q}{\lambda(\rho^q)} \omega^{-2(q-1)\avwf_q} + \omega^3 \frac{\lb(\rho)^q}{\lb(\rho^q)} \omega^{-3(q-1)\avwf_q} = 1
\end{equation}
This equation is perturbative, valid to order $\rho^{2q}$. 
The derivation can be found in the Appendix, which also takes into account higher orders in $\rho$, resulting in a lengthier expression.
Note the similarity of structure between \eqref{eq:wf_av} and the implicit equation obtained for the spectral dimensions (eq \eqref{eq:spec}, see also \cite{Piechon95}).

The resulting theoretical predictions are compared with  numerical results on a finite size system in fig. \eqref{fig:avwf}. The agreement is excellent for all positive values of $q$ compared to the lowest order theory used in \cite{Thiem2013}.

For larger $\rho$, \eqref{eq:wf_av} can be corrected to include higher order terms.
The resulting theoretical prediction (see Appendix) agrees with the numerical computations even for large $\rho$ as shown  in fig. \eqref{fig:avwf}  for the choice $\rho = 0.5$. The reason for this unexpected robustness of our perturbative theory outside its domain of validity is unclear. It supports the idea that the renormalization group picture stemming from the geometrical inflation/deflation property of the Fibonacci chain contains all the fundamental physics determining its electronic properties.

\subsection{The local spectral dimensions and their average}

\begin{figure}[htp]
	\centering
\includegraphics[width=.45\textwidth]{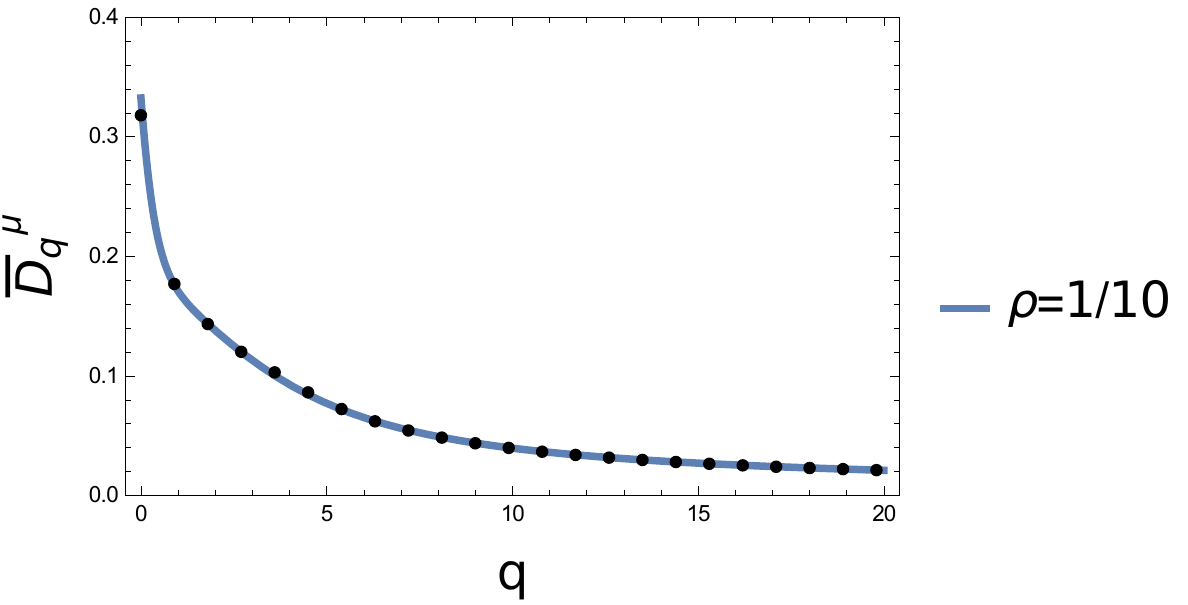}
	\caption{\small{The averaged local spectral dimensions $\avspec_q$ as a function of the multifractal parameter $q$, for $\rho = 0.1$. Dots: numerical results, solid line: theoretical predictions (eq. \eqref{eq:avspec}).}}
	\label{fig:avspec}
\end{figure}

In this section we consider the local density of states, and the associated fractal dimensions.
For a finite-size system, the local density of states  (LDoS) at site $i$ is
\begin{equation}
	\d{\mu_{i}}(E) = \frac{1}{F_n} \sum_{a = 1}^{F_n} \delta(E-E_a) |\psi_{i}(E_a)|^2 \d{E}
\end{equation}
The global density of states is obtained by the sum over all sites of the LDoS.
The local density of states defines $\mu_i$, the local spectral weight at site $i$ associated to the band $a$ of width $\Delta^n_a$:
\begin{equation}
	\mu_i(\Delta^n_a)  = \int_{E \in \text{band }a} \d{\mu_i}(E)
\end{equation}

The local spectral weight at site $i$ sums up all the information about the spectral and wavefunction properties of the Hamiltonian.
To probe the multifractality of the local density of states one defines the partition function
\begin{equation}
	\Gamma_n(q,\tau;i) = \sum_{a} \frac{\left( \mu_i(\Delta^n_a) \right)^q}{(\Delta_a^n)^\tau}
\end{equation}
using which one can compute the local spectral fractal dimensions for individual sites. One can also define a site-averaged gamma function through:
\begin{equation}
	\langle \Gamma_n(q,\tau) \rangle = \frac{1}{F_n} \sum_{i} \Gamma^n(q,\tau;i)
\end{equation}
The site-averaged local spectral dimensions $\avspec_q$, obey the implicit equation:
\begin{equation}
	\label{eq:avspec}
	2 \omega^2 \frac{\lambda(\rho)^q}{\lambda(\rho^q)} z^{(1-q)\avspec_q} +\omega^3  \frac{\lb(\rho)^q}{\lb(\rho^q)} \zb^{(1-q)\avspec_q} = 1
\end{equation}
We compare these theoretical predictions with numerical data in fig. \eqref{fig:avspec}) for $\rho=0.1$ and $q>1$, finding an excellent agreement between the two.
We note, finally, that the theoretical prediction for the Hausdorff dimension $\avspec_0$  given by the equation \eqref{eq:avspec} agrees also well with the numerical result.

\section{\label{sec:discussion}Relations between exponents}
Comparing the relations  \eqref{eq:spec}, \eqref{eq:wf_av} and  \eqref{eq:avspec}  for the global spectral, the average wavefunction and local spectral dimensions respectively, one sees that they bear great similarity. This  suggests that there might be a relation between the three families of exponents. We present below two possible inequalities, to be investigated in future work.

\subsection{A relation between fractal dimensions} 
Consider the case $q=0$. In this case, the local and global spectral dimensions coincide, and the wavefunction dimension is 1, so that we have $\avspec_0 = \avwf_0 D_0$.
A generalization of this relation appears to hold also for all $q>0$, namely $\avspec_q = \avwf_q D_{1+(q-1)\avwf_q}$.
This relation is satisfied numerically for values of $\rho \lesssim 0.2$. 
For larger values of $\rho$, it transforms into the inequality:
\begin{equation}
	\label{eq:relation}
	\avspec_q \geq  \avwf_q D_{1+(q-1)\avwf_q}
\end{equation}

\subsection{An upper bound for the diffusion moments}
Ketzmerick  et al \cite{Ketzmerick1997} obtained a lower bound for the exponent $\sigma_q$, describing the moments of the spreading of a wavepacket: $\sigma_q \geq D_2^\mu / D_2^\psi$.
We propose a new \emph{upper} bound
\begin{equation}
	\sigma_{(1-q)\avwf_q} \leq \frac{D_q^\mu}{D_q^\psi}
\end{equation}
that we derive from the inequality \eqref{eq:relation}, using the relation $\sigma_q = D_{1-q}$ \cite{Piechon96}.
To our knowledge, this is the first upper bound that is proposed for the diffusion exponents.
It is interesting in that it involves the spectral properties (through $D^\mu$), and the wavefunction properties (through $D^\psi$). Work in progress on the study of dynamical correlations on the Fibonacci chain will be reported elsewhere.
 
\section{\label{sec:conclusion}Summary and conclusions}
 In this paper we provide a theoretical description of the spectrum and the wavefunctions of the Fibonacci pure-hopping model in the strong modulation limit, using a perturbative renormalization group scheme. The perturbative approach allows to discuss the structure and labelling of gaps, properties of topological origin and therefore valid for all values of the coupling ratio. We show how using the conumbering basis allows one to characterize wavefunctions conveniently according to their renormalization path. We show that the system has an approximate symmetry, in the perturbative limit, under the exchange of site and energy indices. The leading order expressions for exponents are observed to agree with numerical calculations only for the smallest values of the coupling ratio $\rho$.  We obtain the analytical description of the spectrum and wavefunctions of the Fibonacci chain at next-to-leading order in $\rho$. These expressions show explicitly how the multifractality of wavefunctions appears for larger values of $\rho$. The extended theory is shown to be in very good agreement with numerical results for values of $\rho$ in a wide range from small values all the way upto $\rho$ of order unity. Exponents for the local and global spectral measures for individual states are calculated. Averaged exponents are defined as well, and compared with numerical data. New inequalities relating these exponents, and the diffusion exponent, are proposed and numerically tested.

\emph{ Acknowledgments} We would like to thank J.M. Luck (IPhT, Saclay) and M. Duneau for helpful discussions.

\newpage
\appendix

\section{Appendix}
\label{app:renorm}

In this appendix, we derive the implicit equation the average fractal dimensions of the wavefunctions obey.
In the course of doing so, we find the analytical expression for the renormalization factors $\lambda$ and $\lb$ (equation \eqref{eq:lb}).

the $q$-weight for an energy level $E$ writes
\begin{equation}
	\chi^n(E) = \sum_{i=1}^{F_n} |\psi_i(E)|^{2q}
\end{equation}
We are also going to define the partial sums on atomic (A) or on molecular (M) sites only:
\begin{equation}
	\chi_{A/M}^n(E) = \sum_{i, \text{~at/mol}} |\psi_i(E)|^{2q}
\end{equation}

\emph{Atomic energy levels.}
Let us assume that $E$ is an energy in the atomic cluster at step $n$.
Using the tight-binding equations, we can relate at leading order in $\rho$ the amplitudes on molecular sites to the amplitude on neighboring atomic sites.
Using our Ansatz for the wavefunctions, we relate the amplitude on atomic sites at step $n$ to the amplitude on atomic and molecular sites at step $n-3$.
\begin{equation}
	\chi^n(E) = \lb^q \left( (1+2\rho^{2q}+\rho^{4q})\chi^{n-3}(E') + \rho^{4q}\chi_A^{n-3}(E') \right)
\end{equation}

\emph{Molecular energy levels.}
Using the tight-binding equations, we can relate the amplitudes on atomic sites to the amplitude on neighboring molecular sites. Specifically, we write the $q$-weight on an atomic site $i$ surrounded by two molecular sites $i-1$ and $i+1$ as
\begin{equation}
|\psi_i|^{2q}  = \gamma_q(\rho) \rho^{2q} (|\psi_{i-1}|^{2q}+|\psi_{i+1}|^{2q})
\end{equation}
where we have introduced the function $\gamma_q(\rho)$ that takes into account the possible interferences bewteen the site $i-1$ and $i+1$. We have the constraints $\gamma_0(\rho) = 1/2$, $\gamma_{q \neq 0}(0) = 1$, $\gamma_q(1) = 1/2$. 
We chose for this parameter the form $\gamma_q(\rho) = 1/(1+\rho^{2q})$, which is supported by numerical evidences.
Then, the total $q$-weight on a molecular level writes
\begin{equation}
	\chi^n(E) = \lambda^q \left( (2+\gamma_q \rho^{2q})\chi^{n-2}(E') + \gamma_q \rho^{2q} \chi_A(E') \right)
\end{equation}

\emph{The special case $q=1$.}
When $q=1$, the $q$-norm is independant of $n$. Therefore, we have a closed system of equations for $\lambda$ and $\lb$.
It is straightforward to solve it, and we obtain the expressions given by equation \eqref{eq:lb}.

\emph{The general case: $q$ arbitrary.}
Now the $q$-norm depends on $n$. To obtain a closed system of equations, we take the limit $n \rightarrow \infty$, knowing that the limit behavior of the $q$-norm is $\log \chi^n_q \sim (q-1)\avwf_q \log (1/F_n)$.
After averaging over the energies, we obtain as an implicit equation
\begin{equation}
\label{eq:implicit}
	2 \omega^2 (2\omega^2 I_{MM} + \omega^3 I_{AM}) + \omega^3 ( 2 \omega^2 I_{MA} + \omega^3 I_{AA}) = 1
\end{equation}
with
\begin{align*}
	I_{MM} & = \left( 2+2\rho^{2q}\gamma_q(1 - M(\tau,q)) \right) M(\tau,q)\\
	I_{AM} & =  \left( 2+\rho^{2q}\gamma_q(1 + A(\tau,q)) \right) M(\tau,q)\\
	I_{MA} & =\left( 1 + 2\rho^{2q} + 2\rho^{4q}(1-M(\tau,q)) \right)A(\tau,q)\\
	I_{AA} & =  \left( 1 + 2\rho^{2q} + \rho^{4q}(1+A(\tau,q)) \right)A(\tau,q)
\end{align*}
and the ``molecular'' and ``atomic'' coefficients given by
\begin{align*}
	M(\tau,q) &  = \omega^{-2\tau} \lambda^q \\ 
	A(\tau,q) & = \omega^{-3\tau} \lb^q
\end{align*}
Then, for a given $q$, $\tau = \tau_q^\psi$ is the solution of the implicit equation \eqref{eq:implicit}. The averaged fractal dimensions of the wavefunctions are given by $\avwf_q = \tau_q^\psi/(q-1)$.

\textbf{Perturbative expression in the strong modulation limit:}\\
Neglecting terms of order $\rho^{4q}$ in the above expression, we obtain
\begin{align*}
	I_{MM} \sim I_{AM}& \sim \lambda(\rho)^q/\lambda(\rho^q) \omega^{-2\tau} \\
	I_{MA} \sim I_{AA}&\sim \lb(\rho)^q/\lb(\rho^q) \omega^{-3\tau}.
\end{align*}
Then, from \eqref{eq:implicit} we get the perturbative formula \eqref{eq:wf_av}.

The derivation of the implicit relation for the averaged local spectral dimensions (equation \eqref{eq:avspec}) is in the same lines. The final relation is much simpler because we have dropped the terms of order $\rho^{4q}$. This is in order to be consistent with the results for the spectrum, that are only valid at leading order in $\rho$.
\newpage

\bibliographystyle{unsrt}
\bibliography{bib}
\end{document}